\documentstyle[aps,twocolumn,epsf]{revtex}

\begin{document}

\title{Quasi-localized states in disordered metals and  
non-analyticity of the level curvature distribution function.} 
\date{\today}
\author{ V.\ E.\ Kravtsov$^{1,2}$ and I.\  V.\  Yurkevich$^{1}$.}
\address{$^1$ International Center for Theoretical Physics, P.O. Box
586, 34100 Trieste, Italy\\$^2$
Landau Institute for Theoretical Physics, Kosygina str. 2, 117940 Moscow,
Russia}
\maketitle
\begin{abstract}
It is shown that the quasi-localized states in weakly disordered systems
can lead to the non-analytical distribution 
of level curvatures. In 2D systems the distribution function $P(K)$
has a branching point at $K=0$. In quasi-1D systems the non-analyticity at 
$K=0$
is very weak, and in 3D metals it is absent at all.
Such a behavior confirms the conjecture that the branching at $K=0$
is due to the multi-fractality of wave functions and thus is a generic
feature of all critical eigenstates. The relationsip between the 
branching power and the multi-fractality exponent $\eta(2)$ is derived. 

\end{abstract}
\draft
The problem of energy level statistics in complex quantum systems 
has recently attracted much attention. It is a key problem for the 
description of mesoscopioc fluctuations in small but macroscopic 
disordered electronic systems at low temperatures $^{1}$. It is also of 
great 
fundamental importance, since level statistics is a proper language to 
describe the 
signature of chaos in a quantum system whose classical counterpart is 
chaotic. 

The most of investigations in the field has been aimed to
describe the spectral statistics themselves independently 
of the
nature of wave functions. Nonetheless, it is the space structure of wave 
functions that determines the properties of spectral statistics. In 
systems where the wave functions are extended and essentially structureless 
(e.g. in good metals) the predictions of the random matrix theory (RMT) 
$^{2}$ is proved to work well $^{3}$. 
The corresponding Wigner-Dyson (WD)  spectral statistics 
 is characterized by the level repulsion. On the 
other hand, in the region of localized states, 
energy levels are almost uncorrelated obeying the Poisson
statistics. The spectral statistics at the mobility edge $^{4,5}$ proves 
to be universal but different from both the Poisson and the WD 
statistics. Since the wave functions at the mobility edge are still 
extended, the energy levels repel each other at small energy separations 
$\omega$
of the order of the mean level spacing $\Delta$, the correlations decaying
as a power law $^{5}$ for $\omega\gg\Delta$. 
However, the typical 
fluctuation of the number of levels $\delta N$ in an energy window of the 
width $E=\bar{N}\Delta\gg \Delta$ is much bigger than the RMT prediction 
and behaves in a quasi-Poisson way $^{6,7}$: 
\begin{equation}
\label{s2}
\langle(\delta N)^2 \rangle=\chi \,\bar{N}\gg \ln\bar{N}.
\end{equation}
Such a complex behavior of the critical spectral statistics is
due to the complex statistics of the critical wave functions. A typical 
wave function at the mobility edge can be considered as a superposition 
of peaks with the broad distribution of heights and widths. 
The corresponding quantum-mechanical probability density $|\Psi({\bf 
r})|^2$  is  sparse and is
not negligible only in a small fraction of volume that scales with the 
system size $L$ as $L^{D}$, $D=d-\eta$ being the fractal dimensionality
in the d-dimensional space. However, two wave functions 
 of different energies $|E-E'|\gg\Delta$
overlap well $^{8}$ with 
each 
other since they are strongly correlated in a 
given point: 
\begin{equation}
\label{mf}
\langle|\Psi_{E}({\bf r})|^{p}\,|\Psi_{E'}({\bf r})|^{p}\rangle\propto 
|E-E'|^{-\eta(p)/d}. 
\end{equation}
Such  slowly decaying correlations are typical of the critical wave 
functions. Both in good metal and in the insulator wave functions of 
different energies are locally uncorrelated. The statistics of wave 
functions expressed by Eq.(\ref{mf}) exhibits {\it multi-fractality} $^{9}$, 
since the
fractal dimensionality $D(p)=d-\eta(p)$ depends on the power $p$. 
The main effect of multi-fractality is described by the exponent
$\eta(2)=\eta$. In particular the level compressibility $\chi$ in 
Eq.(\ref{s2}) has been found $^{7}$ to depend on $\eta$ as follows:
\begin{equation}
\label{rid}
\chi=\frac{\eta}{2d}.
\end{equation} 
Multi-fractality of wave functions is typical not only for critical states 
at the mobility edge. Weak multi-fractality holds $^{10-12}$ for 
two-dimensional disordered systems in the regime of weak localization where
the system size $L$ is much less than the localization radius $R_{0}$.

In general, multi-fractality can be considered as a result of 
proliferation of the {\it quasi-localized} states. The latter notion 
$^{13,14}$ means
rare states in disordered systems where $|\Psi({\bf r})|^2$ has a sharp peak
on top of the structureless extended background (in 3D metals) or localized 
states with anomalously small localization radius (for $d\leq  2$).
The existence of such states exhibits itself in the slow relaxation of 
current first discovered in Ref.  $[11]$ using the complicated 
renormalization-group analysis of an extended nonlinear sigma-model in 
$2+\epsilon$ dimensions. 

Very recently the problem of current relaxation in disordered conductors
has been reconsidered $^{14,15}$ by an elegant instanton approximation 
$^{14}$ 
applied to supersymmetric version of the nonlinear sigma-model $^{3}$.
In these works the main result of the previous 
consideration $^{11}$ has been confirmed for 2D systems. However, the 
new method  was able to describe some unknown regimes of current 
relaxation and to set correct limits of validity for the  
regimes found earlier. Later 
the same idea $^{14}$ has been applied $^{12}$ to find directly the 
distribution of 
$|\Psi({\bf r})|^2$ and the distribution of local densities of states 
$^{16}$. Thus the existence of the quasi-localized states has been proved
and the corresponding configuration of the random impurity potential has 
been found $^{14,17}$.

The question we address in this Letter is whether and how the 
quasi-localized 
states and multi-fractality manifest itself in level statistics other than
the level number variance. The natural candidate is the statistics of level
curvatures $K_{n}$ first introduced by Thouless $^{18}$:
\begin{equation}
\label{curv}
K_{n}=\frac{1}{\Delta}\,\left.
\frac{d^2E_{n}(\phi)}{d\phi^2}\right|_{\phi\rightarrow 0}.
\end{equation}
Here $E_{n}(\phi)$ is an energy level depending on the twist of phase in 
the boundary condition $\Psi({\bf r}_{\perp},x=0)=\Psi({\bf 
r}_{\perp},x=L)\,e^{i\phi}$. Such a phase shift arises if an external
magnetic flux threads a sample of the toroidal form.

The main idea of Ref. $[18]$ is that it is possible to distinguish between
the localized and extended states 
studying 
the sensitivity of spectrum to the twist of phase in the boundary 
conditions. This sensitivity is maximal for extended states and  
negligible for the 
localized ones. It is clear that the existence of the quasi-localized 
states should lead to an enhancement in the probability to find a state
with a small value of the random quantity $K_{n}$. Thus the quantity of 
interest is the deviation of the distribution function $P(K)$ from the
form $^{19,20}$ that follows from the random matrix theory.  

This deviation has been recently studied numerically $^{21}$ and 
analytically $^{20,22}$ using an expansion in powers of the inverse 
dimensionless conductance $g^{-1}$. The main
result of the numerical simulations $^{21}$ on the mobility edge of the 3D 
Anderson model is that the deviation
$\delta P(K)=P(K)-P_{WD}(K)\propto |K|^{2-\alpha}$ has a branching point
at $|K|=0$ with $\alpha\approx 0.4$. Although the analytical result
$^{22}$ describes correctly the sign and shape of the deviation
$\delta P(K)$, it is unable to explain the branching at $K=0$
which should be a non-perturbative in $1/g$ effect. It 
has been also conjectured in
$^{21}$ that the branching is due to the multi-fractality of 
critical wave functions and thus the new critical exponent $\alpha$ should
be related with the multi-fractality exponent $\eta$.

In this Letter we will show using the instanton approximation similar to 
that used in $^{14}$ that the level curvature distribution $P(K)$ has, 
indeed, a branching point at $K=0$ in 2D systems under the weak-localization 
condition $R_{0}\gg L$. In 3D metals and quasi-1D systems with $R_{0}\gg L$
the distribution function deviates from the WD form but all its derivatives
are finite at $K=0$. This result favors the statement $^{21}$ that the 
branching in $P(K)$ at $K=0$ and 
multi-fractality are related with each other, since 
for good conductors the multi-fractality is also present only for d=2. 
 
We have found an explicit relationship between $\alpha$ and $\eta$ that 
we believe is valid at the mobility edge near the Anderson transition too:
\begin{equation}
\label{rel1}
\alpha = 3 - 4/\eta,\;\;\;\; (\eta>4/3).
\end{equation}
Another relationship can be found using Eq.(\ref{rid}):
\begin{equation}
\label{al-chi}
\alpha= 3 - \frac{2}{d}\,\chi^{-1},\;\;\;\; (2/d > \chi > 2/3d).
\end{equation}
The first of these equations relates the parametric level 
statistics $P(K)$ with the multi-fractal statistics, Eq.(\ref{mf}), of 
critical wave 
functions. The second one establishes a connection between the parametric 
and the conventional level statistics at the mobility edge. 

The level curvature distribution function 
$P(K)=\langle\delta(K-K_{n})\delta(E-E_{n})\rangle$ 
can be expressed $^{20,22}$ in terms of the  
supersymmetric nonlinear 
sigma-model derived in $^{3}$ from the model of non-interacting electrons 
in a random Gaussian potential: 
\begin{equation}
\label{P-sm}
P(K)=\lim_{\phi\rightarrow 0}\phi^{2}\left[\Re \int {\cal 
D}Q\,A[Q]\,e^{-F[Q;K,\phi]}\right], 
\end{equation} 
where $A[Q]$ is a certain pre-exponent that is irrelevant for the 
present consideration and
\begin{eqnarray}
\label{F}     
F[Q;K,\phi]&=& \frac{\pi}{8}g\int Str\left(\partial  
Q-i\phi {\bf n}\,[Q,\hat{\tau}]\right)^2\,d^d r+\\ \nonumber
&+&i\frac{\pi}{8}K\phi^2 \int Str[\Lambda Q]\,d^d r.
\end{eqnarray}
Here $g=\Delta^{-1}D/L^2$ is the
dimensionless conductance, $D$ is the diffusion coefficient; the unit vector
${\bf n}$ is directed along the $x$-axis.
The field $Q({\bf r})$  is a
$8\times 8$ super-matrix of the certain symmetry $^{3}$ that 
obeys the condition
$Q^2=1$. The symbol
$Str[.]$ stands for the super-trace. 
The matrices 
 $\hat{\tau}= \sigma_{z}(1+\Lambda)/2$ and $\Lambda=diag(1;-1)$, 
describe breaking of time-reversal symmetry and the symmetry between 
retarded and advanced Green's functions, respectively.  Here and below we 
use 
the dimensionless coordinates-ordinates ${\bf r}=\{r_{i}\}$, where 
$-1/2<r_{i}<1/2$.

One can see that the level curvature $K$ enters Eq.(\ref{F})
in the same way as the frequency $\omega$ in the problem of  
long-time current relaxation $^{14}$. 
Therefore from the results of Ref.[14] one would expect the instanton 
approximation to work
well for the Fourier-transform $\tilde{P}(\lambda)=\int dK \,exp[-iK\lambda]$
for $\lambda\gg 1$ rather than for the distribution function $P(K)$ itself.

It is very important that in order to find the distribution function
$P(K)$ or $\tilde{P}(\lambda)$ one should do the limit $\phi\rightarrow 0$.
In this limit a nonzero contribution to the functional
$F[Q;K,\phi\rightarrow 0]$ can arise only from the components  in the 
non-compact boson-boson sector 
of the field $Q$ that can take arbitrary large values.

Another simplification comes from the fact that integration over Grassman 
variables requires an expansion of $\exp\{-F[Q;K,\phi]\}$ in powers of
the Grassman components of the field $Q$. Therefore such components 
can only contribute to the pre-exponent $A$ in the expression
$\tilde{P}(\lambda)=A e^{-S(\lambda)}$, where the effective 
action $S(\lambda)\gg 1$. The main, exponential effect is due to the
conventional variables that parametrize the matrix field $Q$. Thus 
using the Efetov's parametrization $^{3}$ for the orthogonal ensemble we 
have $Q=V^{-1}HV$, where $H$ contains the "non-compact" angles
$\theta_{1,2}({\bf r})\geq 1$ and 
$V$ contains "phase" variables $\varphi({\bf r})\in[0,2\pi]$ and 
$\chi({\bf r})\in[0,2\pi]$. 
All functions must obey the periodic boundary conditions. 

By varying the functional $F[Q;K,\phi]+iK\lambda$ over these functions 
 we find the saddle-point equations to be satisfied by the choice 
$\partial\chi=0$, $\theta_{1}=\theta_{2}=\theta$ and $\varphi({\bf 
r})=\phi v({\bf r})\neq 0$. The equations for the rest two functions read:
\begin{equation}
\label{1}
\partial^2\theta+\phi^2 [\kappa-(\partial v -{\bf n})^2] \sinh\theta=0,
\end{equation}
\begin{equation}
\label{2}
\partial\,[(\partial v-{\bf n}) (\cosh\theta-1)]=0.
\end{equation}
The variation over $K=-2i g\kappa$ leads to the self-consistency equation, 
Eq.(\ref{l}). 

The limit $\phi\rightarrow 0$ is done simply by absorbing $\phi^2$ into
$\theta$. We introduce $\tilde{\theta}=\theta+\ln\phi^2$. 
Then in the limit
$\phi\rightarrow 0$ we have
$\sinh\theta=\cosh\theta=\frac{1}{2}e^{\tilde{\theta}}\,\phi^{-2}\gg 1$
so that $\phi$ drops from Eqs.(\ref{1}),(\ref{2}). 

One can solve Eq.(\ref{2}) to obtain:
\begin{equation}
\label{vsp}
(\partial v -{\bf n})=rot{\bf A}\,e^{-\tilde{\theta}},
\end{equation}
where ${\bf A}$ is an arbitrary vector function. In what follows we 
limit ourselves by considering $rot{\bf A}=
- {\bf n}/N= const$. Integrating Eq.(\ref{vsp}) over space and using 
the periodic boundary conditions for $v({\bf r})$ we obtain the second 
self-consistency equation Eq.(\ref{NN}). 

All equations take a more 
symmetric form if we make a shift $\tilde{\theta}=u-\zeta$, where
$\sinh \zeta =N\kappa^{-1/2}\,(\kappa-N^{-2})/2$:
\begin{equation}
\label{a}
\partial^2 u+\gamma^2 \sinh u=0,
\end{equation}
\begin{equation}
\label{NN}
\frac{1}{N}=\gamma^2 \int e^{-u}\,d^d r,
\end{equation}
\begin{equation}
\label{l}
\lambda=\frac{\pi}{8\gamma^2 N^2}\,\int e^{u}\, d^d r,
\end{equation}
where $\gamma^2=\sqrt{\kappa}/N$.

Solving these equations with the periodic boundary conditions one finds 
$u({\bf r},\lambda)$, $N(\lambda)$
and $\gamma(\lambda)$ that enter the instanton action $S(\lambda)=
\left.F[Q;K,\phi\rightarrow
0]+iK\lambda\right|_{s.p.}$:
\begin{equation}
\label{act}
-\ln \tilde{P}(\lambda)\sim S(\lambda)=\frac{\pi}{4}g\int(\partial 
u)^2 \,d^d \rho + 2 g \gamma^4 N^2 \lambda.
\end{equation}
Eqs.(\ref{a}),(\ref{l}) are analogous to those derived in $^{14,15}$ for the
problem of current relaxation in the orthogonal ensemble. 
The second self-consistency equation (\ref{NN}) is typical for our problem
which is essentially a crossover problem from the orthogonal to the unitary 
ensemble. Another important difference is that we are looking for the
{\it periodic} solutions to the above equations with the period $a=1$
rather than the solutions $^{14}$ that vanish at the boundary of the sample.

Since $\gamma^2\propto K^{1/2}$ we expect the saddle point to correspond to
$\gamma^2 \ll 1$ at $\lambda\rightarrow\infty$. In this case the nonlinear
term in Eq.(\ref{a}) can be approximated by $\gamma^2 \sinh u\approx
\frac{\gamma^2}{2}e^{|u|}\,sign(u)$. Thus we come to the Liouville equation.

Let us first consider the case of a quasi-1D sample. Then the proper 
periodic solution to the Liouville equation takes the form:
\begin{equation}
\label{L1}
e^{u}=4k^2 \gamma^{-2}\,\cosh^{-2} (kx),\;\;\;\;\;\;\;\;\;(|x|<1/4),
\end{equation}
where  $4k^2=\gamma^2\,\cosh^2(k/4)$. In order to obtain the solution $u(x)$ 
for $1/4<|x|<1/2$ one should simply continue the solution, Eq.(\ref{L1}), 
anti-symmetrically about the points $x=\pm 1/4$. Because of such 
anti-symmetricity we have 
\begin{equation}
\label{I}
\int e^{-u}\,d^d r =\int e^{+u}\,d^d r.
\end{equation}
Finally for the case of quasi-1D sample we obtain $N^{-1}=8k\approx 
8\ln(1/\gamma^4)$, $\gamma^4=64\pi k^3/\lambda$, and
\begin{equation}
\label{S1}
\tilde{P}(\lambda)=A \exp\left[-\frac{g_{1}}{2}\ln^2 
\lambda\right],
\end{equation}
where $g_{1}=2\pi g$.
This result is {\it exactly}
of the same form as the long tails of current relaxation $^{14}$ (in the
orthogonal ensemble) if one replaces $t\rightarrow\lambda$.
The region of validity $1\ll
\lambda \ll e^{L/l}$ of Eq.(\ref{S1}) is found from the condition 
of the breakdown of the diffusive regime $^{14,15}$ $|\partial u({\bf 
r})|_{max}=L/l$, where $l$ is the elastic scattering length.

In the 2D case we will look for the continuous periodic solution
to the Liouville equation with the symmetry of a square. It is  
positive 
inside the square  $\Omega$ with vertices at $(0; \pm 1/2)$ and $(\pm 
1/2; 0)$ 
and is negative in the rest half of the sample, being anti-symmetric with 
respect to all sides of the square $\Omega$ and thus obeying  
Eq.(\ref{I}).

The problem of finding such a solution to the Liouville equation can be
solved exactly using the conformal transformations of a unit circle onto
the square $\Omega$. However, for our purposes it is sufficient to know
the solution for $r\ll 1$ where it depends only on $r$. The 
solution that leads to the finite action $S(\lambda)$ is given by:
\begin{equation}
\label{rad}
e^{u(r)}=16 b\,(\gamma^2+b\,r^{2})^{-2},
\end{equation}
where $b=4\pi^4$.
Using this solution we find $N^{-1}=16\pi$, $\gamma^
4=8^3\pi^4/\lambda$, and
\begin{equation}
\label{S2}
\tilde{P}(\lambda)=A\,\lambda^{-2g_{2}}. 
\end{equation}
where $g_{2}=2\pi^2 g$.
Besides the validity condition $1\ll
\lambda\ll (L/l)^4$, this result has again exactly the 
same 
functional form as the long tails of current relaxation $^{14}$ in 2D 
systems. 

In the 3D metal phase, the saddle-point solution for $\tilde{P}(\lambda)$
that depends only on one parameter, the dimensionless conductance, is absent.
The reason is that the 3D Liouville equation does not have a solution that 
corresponds to a finite action $S(\lambda)$ in the limit 
$L/l\rightarrow\infty$. 

It is remarkable  that only in the 2D case the instanton contribution to the
Fourier-transform $\tilde{P}(\lambda)$ of $P(K)$ has a power-law 
asymptotic behavior at $\lambda\gg 1$. In the limit $^{23}$  
$L/l\rightarrow\infty$ this behavior is not restricted from above and the
distribution function $P(K)$ must have a branching point at $K=0$. Namely,
there should exist a set of critical values of 
$g_{2}^{(n)}=\frac{2n+1}{2}=\frac{1}{2},\frac{3}{2},...$ such as for 
$\frac{2n-1}{2}<g_{2}<\frac{2n+1}{2}$ the derivative 
$d^{2n}P(K)/dK^{2n}\propto 
|K|^{-\alpha_{n}}$ at $|K|\rightarrow 0$. The value of $\alpha_{n}$ is 
given by:
\begin{equation}
\label{a-n}
\alpha_{n}=(2n+1)-2g_{2}.
\end{equation}
In particular, for $g_{2}<\frac{1}{2}$, the distribution function $P(K)$
should be divergent at $K\rightarrow 0$. For 
$\frac{1}{2}<g_{2}<\frac{3}{2}$ it should behave like $P(K)-P(0)\propto 
|K|^{2-\alpha}$ where $\alpha=\alpha_{1}=3 - 2 g_{2}$. One can eliminate
the dependence on $g_{2}$ using the expression $^{12}$ for the 
multi-fractality exponent $\eta=\frac{2}{g_{2}}$. Then we arrive at 
the result, Eq.(\ref{rel1}).

In the quasi-1D case the function $\tilde{P}(\lambda)$ falls off fast enough,
and all derivatives of $P(K)$ are finite at $K=0$:
\begin{equation}
\label{der}
P_{2n}\equiv\left.d^{2n}P/dK^{2n}\right|_{K=0} \propto 
\exp[(2n+1)^{2}/2g_{1}]. 
\end{equation}
In this case we have much softer non-analyticity at $K=0$ than that in 2D 
case. However, the function $P(K)$ is still not regular at $K=0$, since
the Taylor series $\sum_{n} P_{2n} K^{2n}/(2n)!$ has  zero radius of 
convergence.

Only in the 3D case the quasi-localized states are not that 
important in the metal phase and the correction $P(K)-P_{WD}(K)$ can be
found perturbatively in $1/g$ as it has been done in Ref.[22].

The above treatment does not allow to take into account localization effects
explicitly. Thus we are not able to {\it derive} the level curvature 
distribution at the mobility edge of 3D disordered systems. However, we 
believe that the distribution function at the mobility edge is similar to 
the one in 2D systems, since both cases are characterized by the 
multi-fractal statistics of wave functions. That is why we believe the 
relationship, Eq.(\ref{rel1}), that we derived for 2D systems in the weak 
localization regime, to hold also in the critical region near the 
Anderson transition.

In summary, we have shown that the quasi-localized states and 
multi-fractality of wave functions can lead to a non-analytical 
distribution of level curvatures $K$. The  
distribution function $P(K)$ is shown to have a branching point at $K=0$ 
for 2D systems and a softer non-analyticity in quasi-1D systems in the 
weak-localization regime. In the 3D metal the distribution function is 
analytical at $K=0$. We argue that the branching in $P(K)$ is typical
for all critical states where the statistics of eigenfunctions exhibit 
multi-fractality. This allows to conjecture a general relationship between
the branching power and the multi-fractality exponent $\eta$. 
{\bf Acknowledgments} We thank C.M.Canali, 
V.I.Falko,
Y.V.Fyodorov, I.V.Lerner, Yu Lu and A.D.Mirlin for stimulating discussions. 
Support from the RFBR-INTAS grant no. 95-0675 (V.E.K.)
is gratefully acknowledged.


\begin{references}
\bibitem{23}
The limit $L/l\rightarrow\infty$ should be taken at a constant $g$
by a simultaneous increase in the
number of transverse channels.
\end{references}
\end{document}